\documentclass[showpacs,showkeys,preprint]{revtex4}
\usepackage{graphicx}

\begin{document}
\title{Search for Majorana Neutrino Signal in $B_c$ Meson Rare Decay}

\author{Shou-Shan Bao$^{a,e}$\footnote{ssbao@sdu.edu.cn},
~~Hong-Lei Li$^b$\footnote{lihl@mail.sdu.edu.cn},~~
Zong-Guo Si$^{a,c}$\footnote{zgsi@sdu.edu.cn},~~
Yi-Bo Yang$^{d,e}$\footnote{yangyb@ihep.ac.cn}}

\affiliation{
$^a$School of Physics, Shandong University, Jinan, 250100, P. R. China\\
$^b$School of Physics and Technology, University of Jinan, Jinan, 250022, P.R.China\\
$^c$Center for High Energy Physics, Peking University, Beijing, 100871, P. R. China\\
$^d$Institute of High Energy Physics, Chinese Academy of Sciences, Beijing, 100049, P. R. China\\
$^e$Kavli Institute for Theoretical Physics China(KITPC), Chinese Academy of Sciences, Beijing, 100190, P. R. China}

\begin{abstract}
We study the $B_c$ meson rare decay in order to search for the Majorana neutrino signal. It is found that the corresponding decay rate is sensitive to the Majorana neutrino mass and mixing angles. The signal of $B_c^\pm\to l_1^\pm l_2^\pm M^\mp $ induced by the Majorana neutrino within the mass region $m_\pi<m_n<m_B$ may be observed at LHCb.
\end{abstract}
\maketitle

%%%%%%%%%%%%%%%%%%%%%%%%%%%%
\section{introduction}
In the Standard Model (SM), the neutrinos are massless since there is no right-hand states. However, non-degenerated neutrino masses provide the most accepted explanation for neutrino experiments \cite{EXP1,EXP2,EXP3,EXP4,EXP5,EXP6,EXP7,EXP8}. This is the favorite evidence for new physics beyond SM. The neutrino masses can be obtained by including the right-hand states, just like the treatment for all other fermions via Yukawa couplings with the Higgs doublet in SM. But unnaturally the hierarchy problem will become more serious due to the small mass of neutrinos. Otherwise as the right-hand neutrinos are SM gauge singlets, the Majorana mass term cannot be ruled out by the gauge invariance. In fact under the help of the Majorana mass term,
it could naturally explain the smallness of the neutrino mass with the so-called see-saw mechanism \cite{Mohapatra:1979ia}. The particular interest in this regard is the question about whether neutrinos are Dirac or Majorana particles. A crucial role to address this question is being played by several experiments looking for the possible existence of the lepton number violation processes.

There have been several attempts to determine the Majorana nature of neutrinos by studying the lepton number violation processes. The experimental observation of such processes may be induced by the Majorana nature of neutrinos.
%It is an avenue to detect $W^{+}W^{+}\to l^{+}l^{+}$ directly on colliders at high energy. Otherwise,
The neutrinoless double beta decays ($0\nu\beta\beta$) in nuclei, $(A,Z)\to (A,Z+2)+2e^-$, have been studied widely. By assuming that $0\nu\beta\beta$ in nuclei are mediated by the exchange of light Majorana neutrinos, the higher precision of present experimental data has been able to set strong constrains on the effective mass, $\langle m_{\alpha\beta}\rangle\equiv \sum_\nu U_{\alpha \nu}U_{\beta \nu}m_{\nu}$, where $\alpha,\beta=e,\mu,\tau$. The upper bound on $\langle m_{ee}\rangle$ is 0.2 eV\cite{Baudis:1999xd}. A global fit\cite{Strumia:2005tc} at 99\% C.L. gives $1.1\times10^{-3}eV\leq
\langle m_{ee}\rangle\leq 4.5\times 10^{-3}eV$(normal hierarchy) or $1.2\times10^{-3}eV\leq\langle m_{ee}\rangle\leq 5.7\times 10^{-3}eV$(inverted hierarchy). From the analysis of atmospheric and solar neutrino oscillation and the tritium beta decay endpoint experiment \cite{Weinheimer:1999tn,Lobashev:1999tp}, the limit is $\langle m_{\mu\mu}\rangle\leq4.4eV$\cite{Barger:1998kz}.   However, it has long been recognized that, even though the experiments are very sensitive, the extraction of the properties of the Majorana neutrinos from nuclear $0\nu\beta\beta$ is a difficult task, because it is reliable only if the nuclear matrix elements for $0\nu\beta\beta$ are calculated precisely.

Another way to detect the Majorana nature is to study the
lepton number violation processes $pp\to l^\pm l^\pm+X$ at
LHC\cite{Chao:2008mq, Si:2008jd,BarShalom:2006bv,Chao:2009ef}. The lepton number violation processes in meson rare decays have also been investigated in refs.\cite{Quintero:2011yh, Ali, Atre:2009rg,Han:2006ip,Atre,Cvetic:2010rw, Claudio,Rodejohann, Zhang:2010um}.
The aim of this work is to
investigate the $B_{C}\to llM$. This signal process can be captured at high intensity experiments such as LHCb\cite{Aaij:2012zr} and future super B factories. The $\Delta L=2$ processes $B_{C}^{\pm}\to l^{\pm}l^{\pm}M^{\mp}$ can
occur via Majorana neutrino exchange, and thus their experimental observation is helpful to
test the Majorana nature of the neutrinos.
%As the see-saw mechanisms suggest both light and heavy neutrinos, to extract the neutrino information through such processes, it's important to distinguish the contribution between the light and heavy neutrinos. In this paper, we study the light, intermediate(which can be on-shell), and heavy neutrino separately.

The paper is organized as follows. The Lagrangian of Majorana neutrinos is introduced in Section \ref{secii}.
The formulas of the $B_c\to l l M$ decays  are obtained in section \ref{seciii}.
In Section \ref{seciv}, we give the numerical results and discussions.  Finally, a short summary is given.

%%%%%%%%%%%%%%%%%%%%%%%%%%%%%%%%%%%%%
\section{Lagrangian related to Majorana neutrino}
\label{secii}
With the same gauge group $SU(2)_{L}\otimes U(1)_{Y}$ in SM, the leptonic content in the simplest extension of the SM includes three generations of left-hand $SU(2)_{L}$ doublets and $n$ right-hand singlets
\begin{equation}
L_{L}=\left(\begin{array}{c}
\nu\\
l
\end{array}\right)_{L},\quad l_{R},\quad N_{R}.
\end{equation}
One can write the general gauge invariant Yukawa terms with Majoranan
mass terms of right-hand neutrinos as
\begin{equation}
-\mathcal{L}_{Y}=f_{l}\bar{L_{L}}\Phi l_{R}+f_{\nu}\bar{L_{L}}\tilde{\Phi}N_{R}+\bar{N_{R}^{c}}M_{R}N_{R}+h.c.
\end{equation}

Therefore, the complete neutrino mass sector is composed of both Dirac mass which is produced via the Yukawa couplings with the Higgs doublet in the SM, and heavy Majorana mass term.
\begin{eqnarray}
-\mathcal{L_{M}} & = & \bar{\nu_{L}}M_{D}N_{R}+\bar{N_{R}^{c}}M_{R}N_{R}+h.c\nonumber \\
 & = & \left(\bar{\nu_{L}},\bar{N_{R}^{c}}\right)\left(\begin{array}{cc}
0 & M_{D}\\
M_{D}^{T} & M_{R}
\end{array}\right)\left(\begin{array}{c}
\nu_{L}^{c}\\
N_{R}
\end{array}\right)+h.c.
\end{eqnarray}
One can find the light Majorana neutrino mass are
\begin{equation}
M_{\nu}\sim-M_{D}M_{R}^{-1}M_{D}^{T},
\end{equation}
 which is called Type I see-saw mechanism. There are other proposals to naturally generate Majorana mass for neutrinos called Type II  or Type III. Generally the mass terms of neutrinos with both Dirac and Majorana terms after gauge symmetry breaking can be expressed as
\begin{eqnarray}
-\mathcal{L} & = & \bar{\nu_{L}}M_{D}N_{R}+\frac{1}{2}\bar{N_{R}^{c}}M_{R}N_{R}+\frac{1}{2}\bar{\nu_{L}}M_{L}\nu_{L}^{c}+h.c.
\end{eqnarray}
To diagnose the mass matrix
\begin{equation}
M=\left(\begin{array}{cc}
M_{L} & M_{D}\\
M_{D}^{T} & M_{R}
\end{array}\right),
\end{equation}
 one need introduce mixing matrix between the gauge and mass eigenstates
\begin{equation}
T=\left(\begin{array}{cc}
U_{3\times3} & V_{3\times n}\\
X_{n\times3} & Y_{n\times n}
\end{array}\right).
\end{equation}
The mixing matrix is unitary, $TT^{\dagger}=T^{\dagger}T=1$.
The states are redefined as follows,
\begin{eqnarray}
\nu_{iL} & \to & U_{ij}\nu_{jL}+V_{ik}N_{kR}^{c},\\
N_{R} & \to & X_{ij}\nu_{jL}^{C}+Y_{ik}N_{kR}.
\end{eqnarray}
As the $M_{N}\gg M_{\nu}$, one has $V\sim X\sim M_{D}M_{N}^{-1}$
and $U^{\dagger}M_{R}+Y^{\dagger}M_{D}\simeq0$.
%
%In the gauge eigenstates, the gauge interaction terms is
%\begin{eqnarray}
%\mathcal{L} & = & \bar{L_{L}}iD\cdot\gamma L_{L}+\bar{N_{R}}i\partial\cdot\gamma N_{R}+\bar{l_{R}}iD\cdot\gamma l_{R}\\
% & = & \bar{\nu_{L}}i\partial\cdot\gamma\nu_{L}+\bar{N_{R}}i\partial\cdot\gamma N_{R}+\bar{l_{L}}i\partial\cdot\gamma l_{L}+\bar{l_{R}}i\partial\cdot\gamma l_{R}\nonumber \\
% &  & +\frac{g}{\sqrt{2}}W_{\mu}^{+}\left(\bar{\nu_{L}}\gamma^{\mu}l_{L}\right)+\cdots+h.c\nonumber
%\end{eqnarray}

%With the mass terms in $\mathcal{L}_{Y}$, the Lagrangian of free
%neutrinos are
%\begin{equation}
%\mathcal{L}_{\nu}=\bar{\nu_{L}}i\partial\cdot\gamma\nu_{L}+\bar{N_{R}}i\partial\cdot\gamma N_{R}-\frac{1}{2}\left(\bar{\nu_{L}},\bar{N_{R}^{c}}\right)\left(\begin{array}{cc}
%M_{L} & M_{D}\\
%M_{D}^{T} & M_{R}
%\end{array}\right)\left(\begin{array}{c}
%\nu_{L}^{c}\\
%N_{R}
%\end{array}\right)+h.c
%\end{equation}

In terms of the mass eigenstates, the gauge interaction Lagrangian
of the charged currents now has the following form,
\begin{eqnarray}
\mathcal
{L}=-\frac{g}{\sqrt{2}}W_\mu^+\left(\sum\limits_{\ell=e}^\tau
\sum\limits_{m=1}^3U_{\ell m}^{\ast }\overline{\nu_m}\gamma^\mu
P_L\ell+\sum\limits_{\ell=e}^\tau\sum\limits_{m^\prime=4}^{3+n}V_{\ell
m^{\prime}}^{\ast}\overline{N_{m^{\prime}}^c} \gamma^\mu
P_L\ell\right)+h.c.
\end{eqnarray}
where $P_L=\frac{1}{2}(1-\gamma_5)$, $\nu_m (m = 1,2,3)$ and
$N_{m^\prime} (m^\prime = 4,\cdots,3+n)$ are the mass eigenstates,
$U_{\ell m}$ is the mixing matrix between the light flavor and light
neutrinos, and $V_{\ell m^\prime}$ is the mixing matrix between the
light flavor and heavy neutrinos.
%
%The external legs are
%\begin{eqnarray}
%\nu|p,\lambda\rangle & = & u^{\lambda}(p)\\
%\bar{\nu}|p,\lambda\rangle & = & \bar{v^{\lambda}}(p)
%\end{eqnarray}
%and
%\begin{eqnarray}
%\langle p,\lambda|\nu & = & v^{\lambda}(p)\\
%\langle p,\lambda|\bar{\nu} & = & u^{\lambda}(p)
%\end{eqnarray}
%
%The vertex are
%\begin{figure}
%\includegraphics{FeynRuleWlN1.eps}\includegraphics{FeynRuleWlN2.eps}\includegraphics{FeynRuleWlnu.eps}
%\caption{The Feynman rules for the charged current vertices in terms of the
%neutrino mass eigenstates.}
%\end{figure}
%
\section{Decay width for  $B_c^{\pm}\to l_1^{\pm}l_2^{\pm}M^{\mp}$}
\label{seciii}
\begin{figure}[htb]
\begin{center}
\begin{minipage}[t]{0.4\textwidth}
\includegraphics[scale=0.3]{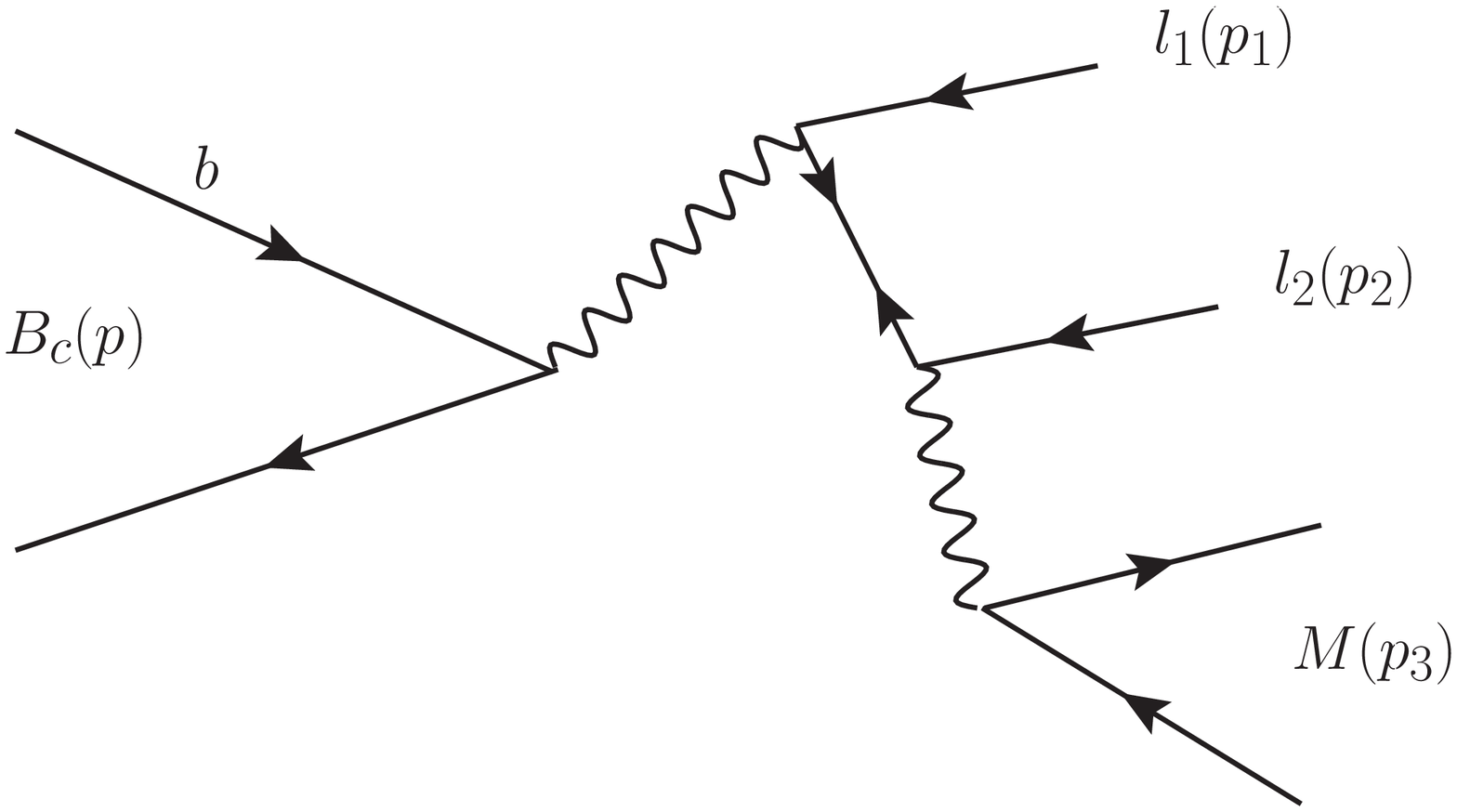}\\
{(a) Annihilation}
\end{minipage}
\begin{minipage}[t]{0.4\textwidth}
\includegraphics[scale=0.3]{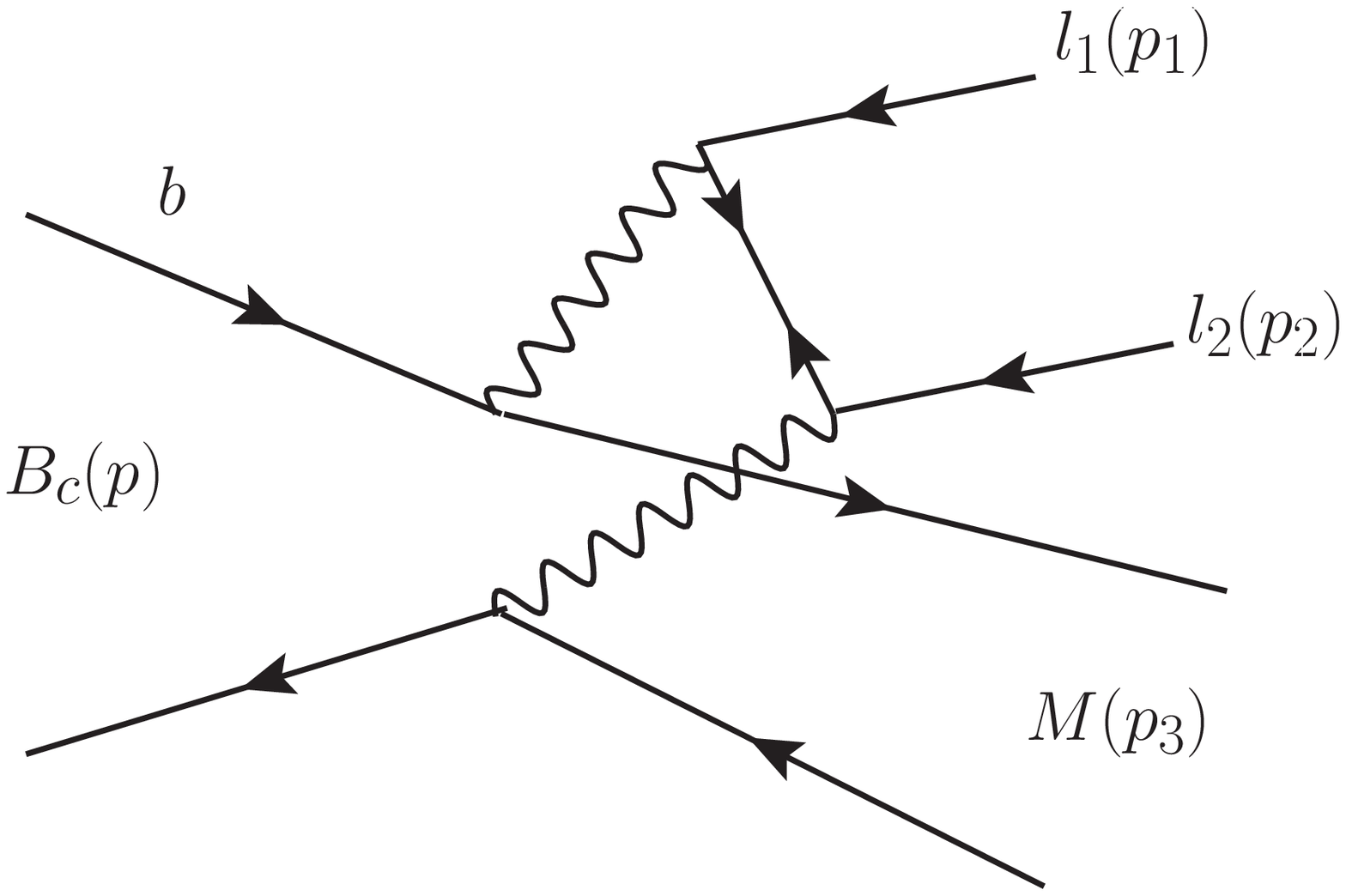}\\
{(b) Emission}
\end{minipage}
\caption{The Feynman diagrams for $B_c\to l^+l^+ M^-$ via Majorana neutrino mediated.}\label{fig:feynmand}
\end{center}
\end{figure}
The Feynman diagrams which contribute to $B_c^{\pm}\to l_1^{\pm}l_2^{\pm}M^{\mp}$ are shown in Fig.(\ref{fig:feynmand}). The first diagram is the annihilation which has been studied widely with the input of meson decay constants. The second one is the emission,  which is considered not enough due to the non-perturbative long distance contributions in the hadronic matrix elements.
%\subsection{The Contribution from the Annihilation Diagrams}
The amplitude of $B_c$ decay to pseudo-scalar contributed by annihilation diagram is expressed as
\begin{equation}
\mathcal{M}_{P}^{a} = -i2G_{F}^{2}V_{B}V_{M}f_{Bc}f_{M}\left(\bar{v_{1}^{C}}p\cdot\gamma p_{3}\cdot\gamma P_{L}v_{2}\right)\sum_\nu S_\nu(p_1)+(1\leftrightarrow2),\label{eq:amplitude}
\end{equation}
where $V_{B}\equiv V_{cb}$, and $V_{M}$ is the corresponding CKM element for final meson. $p$ is the momentum of initial $B_c$ meson. $M_{B}$ is the mass of the meson $B_{C}$. $p_{3}$ is the final meson momentum and $p_1(p_2)$ is the momentum of lepton.  The $f_{B}$ and $f_{M}$ are the decay constants. For a charged or neutral pseudo-scalar meson, the decay constant is defined as
\begin{eqnarray}
&&\langle0|A_{\mu}|P^\pm\rangle=if_{P}p_{\mu},\\
&&\sqrt{2}\langle0|A_{\mu}|P^0\rangle=if_{P}p_{\mu},
\end{eqnarray}
 where the operator $A_\mu$ stands the axial-vector current operator.

The full expression of propagator function $S_\nu$ is
\begin{equation}
  S_\nu(p_1)=\frac{V_{l_1\nu}V_{l_2\nu}m_\nu}{(p_2+p_3)^2-m^2_\nu+im_\nu\Gamma_\nu}.
\end{equation}
The $V_{l_{i}\nu}$ is the lepton mixing matrix element. With the same denotation, the amplitude for $B_c$ decay to vector (or scalar) is
\begin{eqnarray}
\mathcal{M}_{V}^{a} & = & i2G_{F}^{2}M_{V}V_{B}V_{V}f_{B}f_{V}\left(\bar{v_{l1}^{C}}p\cdot\gamma\epsilon_{3}\cdot\gamma P_{L}v_{l_{2}}\right)\sum_\nu S_\nu(p_1)+(1\leftrightarrow2),\label{eq:vamplitude}\\
\mathcal{M}_{S}^{a} & = & -i2G_{F}^{2}V_{B}V_{M}f_{Bc}f_{M}\frac{m_{1}-m_{2}}{M_{M}}\left(\bar{v_{1}^{C}}p\cdot\gamma p_{3}\cdot\gamma P_{L}v_{2}\right)\sum_\nu S_\nu(p_1)+(1\leftrightarrow2).
\end{eqnarray}
In addition to the light neutrinos, the see-saw mechanism predicts very heavy neutrino also. The propagator functions for the two type neutrinos can be approximately expressed as
\begin{equation}
  S(p_1)\simeq\left\{\begin{array}{cc}
                 \frac{V_{l_{1}\nu}V_{l_{2}\nu}m_{\nu}}{(p_2+p_{3})^{2}} & m_{\nu}\ll M_{\pi}, \\
                \frac{V_{l_{1}\nu}V_{l_{2}\nu}}{m_{N}} & m_{N}\gg M_B,
\end{array}\right.
\end{equation}
 where we still use $\nu$ to denote for the light neutrino, and $N$ for the one heavier than $m_B$.

 In the meantime, there is another possibility that the neutrino mass is between the $M_B$ and the final meson mass which is denoted with $n$ ($m_{\pi}<m_n<m_B$).
Neutrinos with such mass has been strongly constrained by direct search and  cosmological observations\cite{Merle:2006du,Seljak:2006bg,Kusenko:2009up,Dunkley:2008ie} and must be sterile. The process of $B_c^\pm\to l_1^\pm l_2^\pm M^\mp$ is dominated by the annihilation diagram as the intermediated neutrino in s-channel can be on-shell.
At this case the narrow width approximation
 \begin{equation}
  \lim_{\Gamma\to 0} \frac{1}{(q^2-m^2)^2+m^2\Gamma^2}=\frac{\pi}{m\Gamma}\delta(q^2-m^2),
 \end{equation}
can be applied. As the $l^{\pm}+X$ are the dominated decay channels, the total decay width of the neutrino can be expressed as\cite{Cvetic:2010rw},
\begin{equation}
\Gamma_{n}=2\sum_{l}|V_{ln}|^{2}\left(\frac{m_{n}}{m_{\tau}}\right)^{5}\times\Gamma_{\tau},
\end{equation}
where the $m_{\tau}$, $\Gamma_{\tau}$ are the mass and total width
of the tau-lepton. With the total decay width and narrow width approximation, one can get the decay width of $B_c^\pm\to l_1^\pm l_2^\pm\pi^\mp$
\begin{equation}
\Gamma(l_1l_2\pi)=\frac{G_F^4|V_{B}V_{\pi}|^{2}f_{B}^{2}f_{\pi}^{2}}{128\pi^{2}}\frac{|V_{l_1n}V_{l_2n}|^{2}}{\sum_{l}|V_{ln}|^{2}}\frac{m_{B}m_{\tau}^{5}}{2\Gamma_{\tau}}\left(1-\frac{m_\pi^{2}}{m_{n}^{2}}\right)^{2}\left(1-\frac{m_{n}^{2}}{m_{B}^{2}}\right)^{2}.
\end{equation}

In summary, the square of annihilation amplitudes can be written as
\begin{eqnarray}
|\mathcal{M}^a_{P}|^{2} & = & 8\left(G_{F}^{2}V_{B}V_{M}f_{Bc}f_{M}\right)^{2}\left(p_{1}\cdot p_{2} \right)F(p_1,p_2)+(p_1\leftrightarrow p_2),\\
|\mathcal{M}^a_{S}|^{2} & = & 8\left(G_{F}^{2}V_{B}V_{M}f_{Bc}f_{M}\right)^{2}\left(\frac{m_{1}-m_{2}}{M_{s}}\right)^{2}\left(p_{1}\cdot p_{2}\right)F(p_1,p_2)+(p_1\leftrightarrow p_2),\\
|\mathcal{M}^a_{V}|^{2} & = & 4\left(G_{F}^{2}V_{B}V_{M}f_{Bc}f_{M}\right)^{2}M_{V}^{2}\frac{\mathcal{V}(p_{1},p_{2})}{(p-p_{1})^{4}}F(p_1,p_2)+(p_1\leftrightarrow p_2),
\end{eqnarray}
where the function ${F}(p_1,p_2)$ and $\mathcal{V}(p_1,p_2)$ are defined as,
\begin{eqnarray}
{F}(p_1,p_2)&=&\left\{\begin{array}{l l}
          |\sum_\nu V_{l_{1}\nu}V_{l_{2}\nu} m_{\nu}|^2, & m_{\nu}\ll m_{\pi}, \\
         |V_{l_1n}V_{l_2n}|^2\frac{\pi m^5_n}{\Gamma}\delta((p-p_1)^2-m_n^2), & \textrm{on shell},\\
         |V_{l_1N}V_{l_2N}|^2\frac{(p-p_1)^4}{m^2_N}, & m_{N}\gg m_{B},\end{array}\right.\\
\mathcal{V}(p_1,p_2)% & = & 2\frac{p_{1}\cdot p_{2}}{M_V^{2}}(p_{1}-p_{3})^{4}-4M_V^{2}p_{1}\cdot p_{2}+8p_{1}\cdot p_{3}p_{2}\cdot p_{3}\nonumber\\
 & = & 2p_{1}\cdot p_{2}\left(4p_{1}\cdot p_{3}-M_V^{2}+\frac{4(p_{1}\cdot p_{3})^{2}}{M_V^{2}}\right)+8p_{1}\cdot p_{3}p_{2}\cdot p_{3}.
\end{eqnarray}

In addition to the annihilation diagrams, we also consider the emission diagrams. We apply the light-cone function of mesons to calculate the hadronic amplitude. The non-perturbative effect from long-distance interaction is factorized into the light-cone function, and the leptonic number violation effect is caused by the short-distance interaction which can be calculated perturbatively.
The light-cone distribution amplitude is defined as\cite{Beneke:2000wa}
\begin{equation}
\mathcal{M}_{\beta\alpha}^{M}(k) = \int\frac{d^{4}k}{(2\pi)^{4}}e^{-ik\cdot x}\langle0|\bar{q}_{\alpha}(x)q_{\beta}(0)|M\rangle,
\end{equation}
and parameterized with the twist wave functions as
\begin{eqnarray}
\mathcal{\mathcal{M}^{B}} & = & -\frac{if_{B}}{4}\left\{ (p\!\!\!/+m_{B})\gamma^{5}\phi_{B}(u)\right\}, \\
\mathcal{M}^{\pi} & = & -\frac{if_{\pi}}{4}\left\{ p\!\!\!/\gamma^{5}\phi(u)-\mu_{\pi}\gamma^{5}\left(\phi_{P}(u)-i\sigma_{\mu\nu}n_{-}^{\mu}v^{\nu}\frac{\phi_{\sigma}^{'}(u)}{6}+i\sigma_{\mu\nu}p^{\mu}\frac{\phi_{\sigma}(u)}{6}\frac{\partial}{\partial k_{T\nu}}\right)\right\}.
\end{eqnarray}

For the wave function of $B_c$,
we take the following form in the numerical calculations\cite{Cheng:2001aa,Cheng:2000hv},
\begin{equation}
\phi_{B}(x) = N_{B}x^{2}(1-x)^{2}\exp\left[-\frac{1}{2}\left(\frac{xm_{B}}{\omega_{B}}\right)^{2}\right],
\end{equation}
where $\omega_B$ is the shape parameter, and $N_B$ is the normalization constant.
The general expression of twist-2 wave function for pion is
\begin{equation}
\phi_{\pi}(x)  = 6x(1-x)\left(1+\sum_{n=1}a_{n}C_{n}^{3/2}(2x-1)\right),
\end{equation}
where $C_{n}(x)$ is the Gegenbauer polynomial. In this work, the higher twist contributions are not considered.

For the wave functions of vector mesons, involving one longitudinal(L) and two transverse(T)
polarizations, can be expressed as\cite{Liu:2009qa},
\begin{eqnarray}
\mathcal{M}^V_L&=&\frac{1}{\sqrt{6}}\left\{M_V\epsilon\!\!\!/^{*L}_V\phi_V(u)+\epsilon\!\!\!/^{*L}_Vp\!\!\!/\phi_V^t(u)+M_V\phi_V^s(u)\right\},\\
\mathcal{M}^T_V&=&\frac{1}{\sqrt{6}}\left\{M_V\epsilon\!\!\!/^{*T}_V\phi^v_V(u)+\epsilon\!\!\!/^{*T}_Vp\!\!\!/\phi_V^T(u)+M_Vi\epsilon_{\mu\nu\rho\sigma}\gamma^5\gamma^\mu\epsilon_V^{*\nu}n^\rho v^\sigma\phi_V^a(u)\right\}.
\end{eqnarray}
where $\epsilon^{L(T)}_V$ denotes the longitudinal(transverse) polarization vector. The distribution amplitudes $\phi^{(t,s)}_V$ and $\phi_V^{v(T,a)}$ can be parameterized as
\begin{eqnarray}
&&  \phi_V(x)=\frac{3f_V}{\sqrt{6}}x(1-x)\left[1+a^\|_{1V}C_1^{3/2}(2x-1)+a_{2V}^|C_2^{3/2}(2x-1)\right],\\
&&  \phi_V^T(x)=\frac{3f_V^T}{\sqrt{6}}x(1-x)\left[1+a_{1V}^\bot C_1^{3/2}(2x-1)+a_{2V}^\bot C_2^{3/2}(2x-1)\right],\\
&&  \phi_V^t(x)=\frac{3f_V^T}{2\sqrt{6}}(2x-1)^2,\quad\qquad\phi_V^s(x)=\frac{3f_V^T}{2\sqrt{6}}(1-2x),\\
&&  \phi_V^v(x)=\frac{3f_V}{8\sqrt{6}}(1+(2x-1)^2),\quad\phi_V^a(x)=\frac{3f_V}{4\sqrt{6}}(2x-1).
\end{eqnarray}

The amplitude of the contribution from emission Feynman diagrams can be written as
\begin{equation}
\mathcal{M}^{E}= \int dx dy~ 2G_{F}^{2}V_{ub}V_{cd}\left[\frac{\langle m_{l_{1}l_2}\rangle}{q^{2}+i\epsilon}+\sum_{N}\frac{V_{l_{1}N}V_{l_{2}N}}{m_{N}}\right]W^{\mu\nu}L_{\mu\nu}+(1\leftrightarrow2),\label{eq:emision}
\end{equation}
where $q=xp_{B}-yp_{3}-p_{2}$. $x$ and $y$ are the momentum fractions. The first term in Eq.(\ref{eq:emision}) is the contribution of light neutrinos while the second term is that of heavy neutrinos. The leptonic and hadronic tensors are
\begin{eqnarray}
L_{\mu\nu}&=&\left(\bar{v_{l1}^{C}}\gamma_{\mu}\gamma_{\nu}P_{L}v_{l_{2}}\right),\\
W^{\mu\nu} & = & \langle\pi|\left(\bar{b}\gamma^{\mu}(1-\gamma^{5})u\right)\left(\bar{d}\gamma^{\nu}(1-\gamma^{5})c\right)|B_{c}\rangle\nonumber\\
& = & \frac{f_{B}f_{\pi}}{2N_{C}}\left(p^{\mu}p_{\pi}^{\nu}+p^{\nu}p_{\pi}^{\mu}-p\cdot p_{\pi}g^{\mu\nu}\right)\phi_{B}(x)\phi_{\pi}(y),
\end{eqnarray}
where $N_C$ is the color factor. One can get the emission amplitude as
%
%\begin{eqnarray}
%W^{\mu\nu}L_{\mu\nu} & = & -\frac{f_{B}f_{\pi}p\cdot p_{\pi}}{N_{C}}\left(\bar{v_{l1}^{C}}P_{L}v_{l_{2}}\right)\phi_{B}\phi_{\pi}
%\end{eqnarray}
\begin{eqnarray}
\mathcal{M}^{E} & = &\frac{2G_{F}^{2}V_{ub}V_{cd}f_{B}f_{\pi}p\cdot p_{\pi}\phi_{B}\phi_{\pi}}{N_{C}}~\left(\bar{v_{l1}^{C}}P_{L}v_{l_{2}}\right)\nonumber\\
&&\times\left[\int dxdy \frac{\langle m_{l_1l_2}\rangle}{q^{2}+i\epsilon}+\sum_{n}\frac{V_{l_{1}N}V_{l_{2}N}}{m_{N}}\right]+(1\leftrightarrow2).\label{eq:emission}
\end{eqnarray}
Comparing Eq.(\ref{eq:emission}) with Eq.(\ref{eq:amplitude}), one can notice that the contribution of heavy neutrino in emission diagrams is similar to that in annihilation diagrams but suppressed by the color factor $N_C$, while in some channel also suppressed by the CKM matrix element. Thus the emission diagrams can be considered only for light neutrinos.

%%%%%%%%%%%%%%%%%%%%%%%%%%
\section{Numerical Results and Discussion}
\label{seciv}
%In the last section, the formulas of the contributions in different cases have been given. In this section, we discuss the results numerically.
To compare the contribution from the annihilation diagram with that from emission one, we study them
separately. The numerical values of the mesons' decay constants and mass \cite{Nakamura:2010zzi, Cvetic:2004qg,Wang:2007fs,Ball:2006eu} are listed in Table.(\ref{tb:input_p},\ref{tb:input_v}).
\begin{table}
\begin{tabular}{|c|c|c|c|c|c|c|}
\hline
P & $\pi^{-}(\bar{u}d)$ & $K^{-}(\bar{u}s)$ & $D^{-}(\bar{c}d)$ & $D_{s}^{-}(\bar{c}s)$ & $B^{-}(\bar{u}b)$ & $B_{c}^{+}(\bar{c}b)$\\
\hline
\hline
$f_{P}$ & $130.41$ & $156.1$ & $206$ & $257.5$ & $193$ & $322$\\
\hline
$M_{P}$ & $139.57$ & $493.677$ & $1869.60$ & $1968.47$ & $5279.17$ & $6.277$\\
\hline
\end{tabular}
\caption{The input parameters for pseudo-scalars}\label{tb:input_p}
\end{table}
\begin{table}
\begin{tabular}{|c|c|c|c|c|}
\hline
V & $\rho^{-}(\bar{u}d)$ & $K^{*-}(\bar{u}s)$ & $D^{*-}(\bar{c}d)$ & $D_{s}^{*-}(\bar{c}s)$ \\
\hline
\hline
$f_{V}$ & 216 & 220 & 240 & 272 \\
\hline
$m_{V}$ & 770 & 891.66 & 2010.22 & 2112.3 \\
\hline
\end{tabular}
\caption{The input parameters for vector mesons.}\label{tb:input_v}
\end{table}
For the light neutrinos($m_\nu\ll m_\pi$), we take the effective mass of the light neutrinos $\langle m_{l_{1}l_{2}}\rangle=|\sum_{\nu}V_{l_{1}\nu}V_{l_{2}\nu}m_{\nu}|$ as input. The contribution from the annihilation diagrams is found to be
\begin{eqnarray}
Br(B^\pm_{c}\to l_1^{\pm}l_2^{\pm}\pi^{\mp}) & \simeq & 1.4\times10^{-33}\left(\frac{\langle m_{l_{1}l_{2}}\rangle}{1eV}\right)^{2}.
%Br(B_{c}\to l^{+}l^{+}\rho^{-}) & \simeq & 6.0\times10^{-32}\left(\frac{\langle m_{l_{1}l_{2}}\rangle}{1eV}\right)^{2}.%\\
%Br(B_{c}\to l^{+}l^{+}a^{-}) & \simeq & 1.45\times10^{-33}\left(\frac{\langle m_{l_{1}l_{2}}\rangle}{1eV}\right)^{2}
\end{eqnarray}
One can find the  results for other channels in Table.(\ref{tb:light}) with $\langle m_{l_1l_2}\rangle=1eV$.
\begin{table}[htb]
\begin{center}
\begin{tabular}{|c|c|c|c|c|}
\hline
P & $\pi^{-}(\bar{u}d)$ & $K^{-}(\bar{u}s)$ & $D^{-}(\bar{c}d)$ & $D_{s}^{-}(\bar{c}s)$\\
\hline
Br & $1.4\times10^{-33}$ & $9.6\times10^{-35}$ & $6.5\times10^{-35}$ & $1.8\times10^{-33}$\\
\hline
V & $\rho^{-}(\bar{u}d)$ & $K^{*-}(\bar{u}s)$ & $D^{*-}(\bar{c}d)$ & $D_{s}^{*-}(\bar{c}s)$\\
\hline
Br & $6.0\times10^{-32}$ & $2.3\times10^{-33}$ & $3.0\times10^{-34}$ & $6.4\times10^{-33}$\\
\hline
\end{tabular}
\caption{The branch ratio of $B_c^+\to l_1^+l_2^+M^-$  with $\langle m_{l_1l_2}\rangle=1eV$ from annihilation.}\label{tb:light}
\end{center}
\end{table}

The results indicate that the branch ratio of $B^\pm_{c}\to l_1^{\pm}l_2^{\pm}P^{\mp}$
is about $10^{-33}\times|V_{P}|^{2}\times\left(\frac{\langle m_{l_1l_2}\rangle}{1eV}\right)^{2}$.
Due to the suppression of the CKM elements $V_{us}$,$V_{cd}$, the branch
ratio to $K(K^*)$ and $D(D^*)$ are about one order smaller than $\pi(\rho)$ and
$D_s(D_s^*)$. As the possible relevant phase between $V_{l_{1}\nu}$and $V_{l_{2}\nu}$,
the $\langle m_{l_{1}\nu l_{2}\nu}\rangle$ may be very small which indicates invisible contributions.

\begin{figure}[htb]
\begin{center}
\includegraphics[scale=0.8]{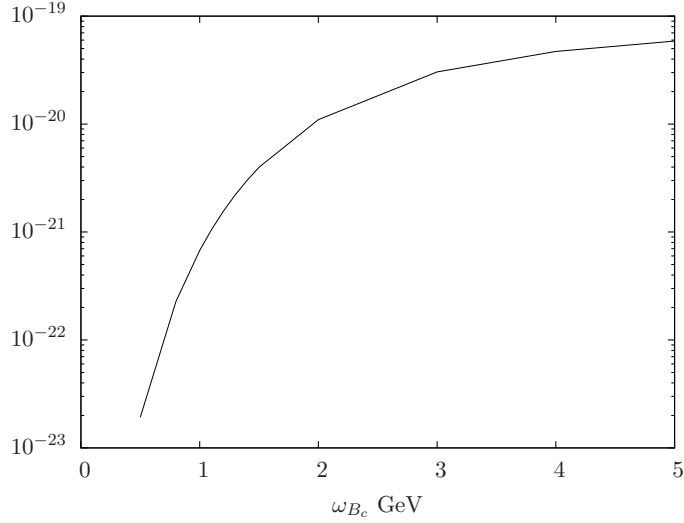}
\caption{The branch ratio of $B_c^\pm\to l^\pm l^\pm \pi^\mp$ as a function of the shape parameter $\omega_{Bc}$.}\label{fig:omega}
\end{center}
\end{figure}

In addition to the annihilation diagrams, the contribution from emission diagram is depend on the non-perturbative parameter $\omega_B$.  For $\omega_B=1.0GeV$,  the branch ratio of $B^\pm_{c}\to l^{\pm}l^{\pm}\pi^{\mp}$ contributed by emission diagram is \begin{equation}
Br(B^\pm_{c}\to l_1^{\pm}l_2^{\pm}\pi^{\mp})=6.75\times10^{-22}\left(\frac{\langle m_{l_1l_2}\rangle}{1eV}\right)^2.
\end{equation}
In Fig.(\ref{fig:omega}) the result as a function of $\omega_B$ is shown.
 One can notice the emission diagram is dominated comparing to the annihilation diagrams for light Majorana neutrinos. However it is still below the experimental bounds.

Now we consider the effect of heavy neutrino($m_{N}\gg m_{B}$).
The nature explaining to the smallness of neutrino masses in the see-saw mechanism need $m_N\gg\Lambda_{EW}$ where $\Lambda_{EW}$ is the Electro-Weak scale. The mixing element $V_{lN}$ is also very small. One can image that $V_{1N}V_{2N}/m_N\sim1/\Lambda_{GUT}$. If $V_{1N}V_{2N}/m_N=10^{-16}GeV^{-1}$ is taken, one can get the annihilation result
\begin{eqnarray}
Br(B_c^\pm\to l_1^{\pm}l_2^{\pm}\pi^\mp) & \simeq & 3.3\times10^{-45}\left(\frac{|V_{l1N}V_{l_{2}N}|/m_{N}}{10^{-16}GeV^{-1}}\right)^{2}.
%Br(l^+l^+\rho^-)&\simeq& 6.1\times 10^{-45}\left(\frac{|V_{l1N}V_{l_{2}N}|/m_{N}}{10^{-16}GeV^{-1}}\right)^{2}.
\end{eqnarray}
 The results related to other channels with $V_{1N}V_{2N}/m_N=10^{-16}GeV^{-1}$ are listed in Table.(\ref{tb:heavy}).
\begin{table}[htb]
\begin{center}
\begin{tabular}{|c|c|c|c|c|}
\hline
P & $\pi^{-}(\bar{u}d)$ & $K^{-}(\bar{u}s)$ & $D^{-}(\bar{c}d)$ & $D_{s}^{-}(\bar{c}s)$\\
\hline
Br & $3.3\times10^{-45}$ & $2.1\times10^{-46}$ & $9.6\times10^{-47}$ & $2.6\times10^{-45}$\\
\hline
V & $\rho^{-}(\bar{u}d)$ & $K^{*-}(\bar{u}s)$ & $D^{*-}(\bar{c}d)$ & $D_{s}^{*-}(\bar{c}s)$\\
\hline
Br & $6.1\times10^{-45}$ & $3.4\times10^{-45}$ & $3.5\times10^{-46}$ & $6.6\times10^{-45}$\\
\hline
\end{tabular}
\caption{The branch ratio of $B_c^\pm\to l_1^{\pm}l_2^{\pm}M^\mp$ with ${V_{1N}V_{2N}/m_N}={10^{-16}}GeV^{-1}$. Only the annihilation diagrams are considered, since the emission diagrams are suppressed.}\label{tb:heavy}
\end{center}
\end{table}
One can notice that the contribution of such massive neutrinos to $B_c^\pm\to l_1^{\pm}l_2^{\pm}M^\mp$ can be omitted comparing with contribution of light neutrinos. As pointed in last section, the contribution of massive neutrinos in emission diagrams is suppressed and more less.

%\subsection{The Contribution from Emission Diagrams}
%
%\begin{table}
%\begin{center}
%\begin{tabular}{|c|c|c|c|c|c|c|c|c|c|c|c|c|}
%\hline
%$\omega_{Bc}$GeV & 0.5 & 0.8 & 1.0 & 1.1 & 1.2 & 1.3 & 1.4 & 1.5 & 2.0 & 3.0 & 4.0 & 5.0\\
%\hline
%Br$(10^{-22})$ & $0.192$ & $2.27$ & $6.75$ & $10.5$ & $15.6$ & $22.1$ & $30.1$ & $39.8$ & $110$ & $304$ & $471$ & $588$\\
%\hline
%\end{tabular}
%\caption{The Branch ratio with different $\omega_{Bc}$.}
%\end{center}
%\end{table}

\begin{table}
\begin{center}
\begin{tabular}{c|c|c|c}
\hline
                        &$m_{\nu}<<m_{\pi}$  &$m_n\sim m_B^*$   &$m_n>>m_B$\\
\hline
\textrm{Emission}       &$\sim10^{-21}$   & $\sim10^{-21}$ &    $<10^{-46}$  \\
\hline
\textrm{Annihilation}   &$\sim10^{-32}$   & $<10^{-4}$ &    $\sim10^{-45}$  \\
\hline
\end{tabular}
\caption{The comparing of the contribution from emission and annihilation diagrams.}\label{tb:compare}
\end{center}
\end{table}

\begin{figure}[hbt]
\includegraphics[scale=0.25]{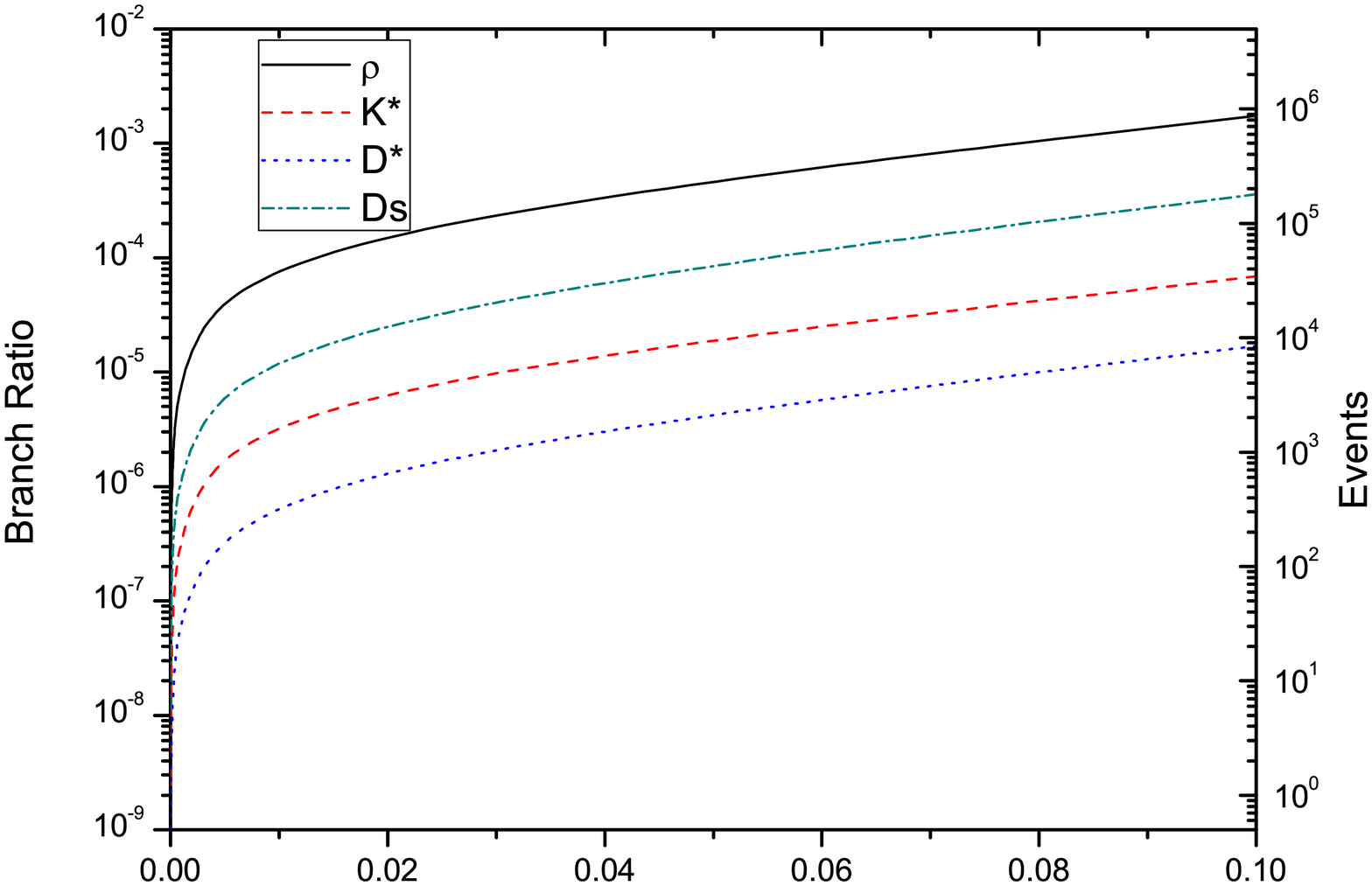}
\includegraphics[scale=0.25]{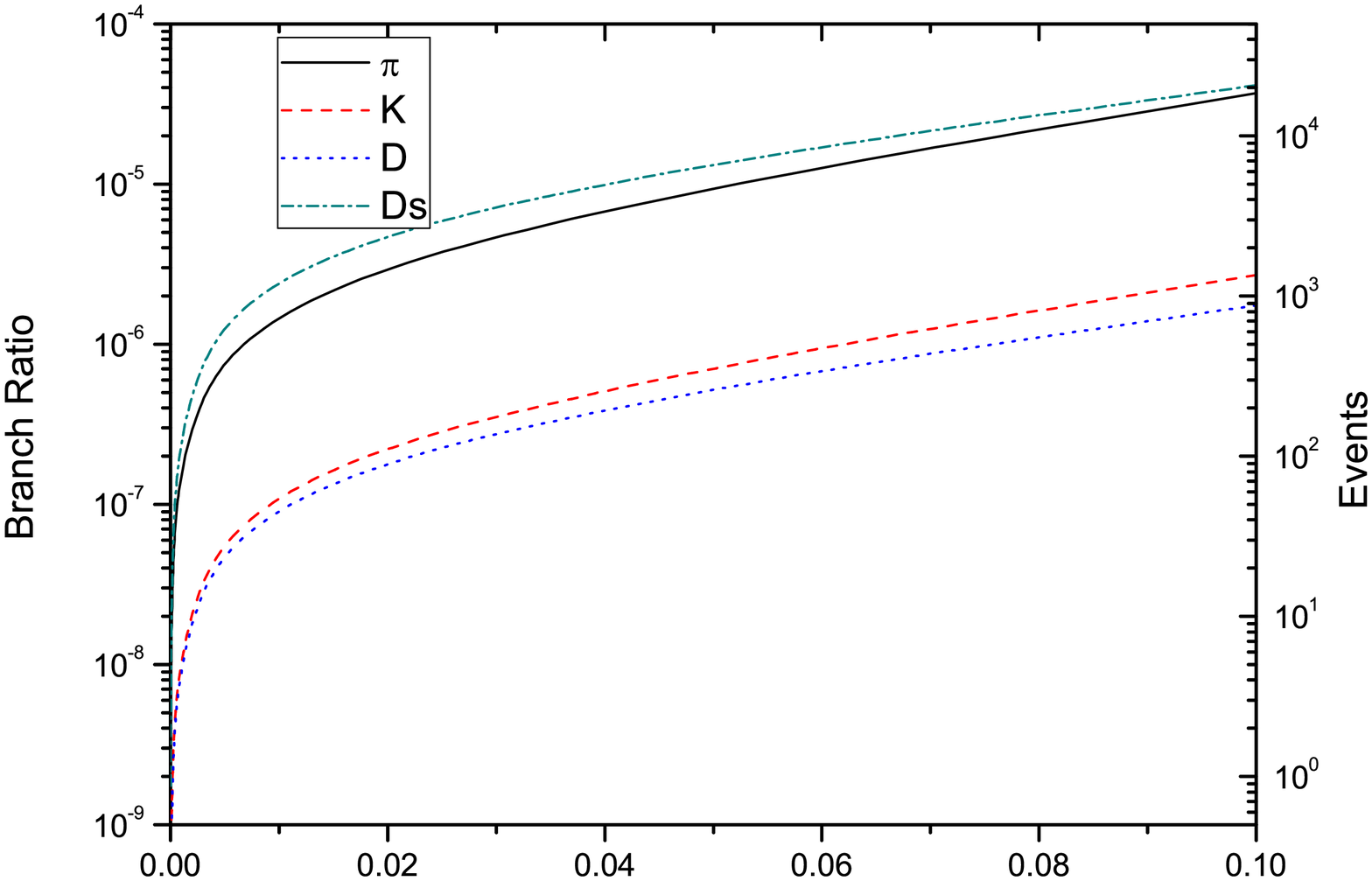}
\caption{The branch ratio of $B_c$ and the corresponding event number on LHCb as functions of $\frac{|V_{l_1n}V_{l_2n}|^2}{\sum_l |V_{ln}|^2}$ with $m_n=m_B/2$.}\label{fig:threshold}
\end{figure}

%\subsection{The contribution of the intermediate neutrino $m_\pi<m_n<m_B$}
%
%Although the light neutrino mass has been strongly constrained by direct search and  cosmological observations\cite{Seljak:2006bg,Kusenko:2009up,Dunkley:2008ie}, we consider a neutrino with mass $m_{\pi}<m_{n}<m_{B}$. Majorana neutrino with such mass must be sterile and constrained strongly by experiments. We study the possible constrains through the $\Delta L=2$ semi-leptonic decay on LHCb. The process of $B_c\to M l l$ is dominated by the annihilation diagram as the intermediated neutrino in s-channel can be on-shell.
%At this case the narrow width approximation
% \begin{equation}
%  \lim_{\Gamma\to 0} \frac{1}{(q^2-m^2)^2+m^2\Gamma^2}=\frac{\pi}{m\Gamma}\delta(q^2-m^2),
% \end{equation}
% can be applied.
%As the $l^{\pm}+X$ are the dominated decay channels, the total decay width of the neutrino can be expressed as\cite{Cvetic:2010rw},
%\begin{equation}
%\Gamma_{n}=2\sum_{l}|V_{ln}|^{2}\left(\frac{m_{n}}{m_{\tau}}\right)^{5}\times\Gamma_{\tau},
%\end{equation}
%where the $m_{\tau}$, $\Gamma_{\tau}$ are the mass and total width
%of the tau-lepton. With the total decay width and narrow width approximation, one can get the decay width of $B_c\to ll\pi$
%\begin{equation}
%\Gamma(l_1l_2\pi)=\frac{G_F^4|V_{B}V_{\pi}|^{2}f_{B}^{2}f_{\pi}^{2}}{128\pi^{2}}\frac{|V_{l_1n}V_{l_2n}|^{2}}{\sum_{l}|V_{ln}|^{2}}\frac{m_{B}m_{\tau}^{5}}{2\Gamma_{\tau}}\left(1-\frac{m_\pi^{2}}{m_{n}^{2}}\right)^{2}\left(1-\frac{m_{n}^{2}}{m_{B}^{2}}\right)^{2}.
%\end{equation}
At last we discuss the sterile neutrinos with mass $m_\pi<m_n<m_B$. We take $m_{\tau}=1.77GeV$, $\Gamma_{\tau}=2.3\times10^{-12}GeV$
and $m_{n}=m_{B}/2$.  The intermediated neutrino in annihilation diagram can be on shell which will enhance such process. The emission diagram contribution has no enhancement which can be omitted.
We list the approximate contribution from emission and annihilation diagrams in Table. \ref{tb:compare}. With such inputs, the  numerical result of branch ratio for $B_c^\pm\to l_1^{\pm}l_2^{\pm}\pi^\mp$ is
\begin{equation}
Br(B_c^\pm\to l_1^{\pm}l_2^{\pm}\pi^\mp)\simeq4.39\times10^{-4}\times\frac{|V_{l_{1}n}V_{l_{2}n}|^{2}}{\sum_{l}|V_{ln}|^{2}}.
\end{equation}
The $\frac{|V_{l_{1}n}V_{l_{2}n}|^{2}}{\sum_{l}|V_{ln}|^{2}}$ is a parameter relating with mixing between the leptons. The branch rations as a function of the parameter are shown in Fig.(\ref{fig:threshold}). One can  notice that the rare decay $B_c^\pm\to l^\pm l^\pm \pi^\mp$ may be detected on LHCb at this case.  It is expected to study the property of the Majorana neutrinos indirectly through such processes.

The production of $B_c$ at LHCb have been studied widely. The gluon-gluon fusion subprocess $gg\to B_c+X$ is the dominated production channel and much larger than the quark-antiquark annihilation subprocess $q\bar{q}\to B_c+X$. The magnitude of the color-octet components may be estimated with the non-relativistic QCD and the contribution can be neglected comparing with color-singlet components.

\begin{figure}[hbt]
\begin{center}
  \includegraphics{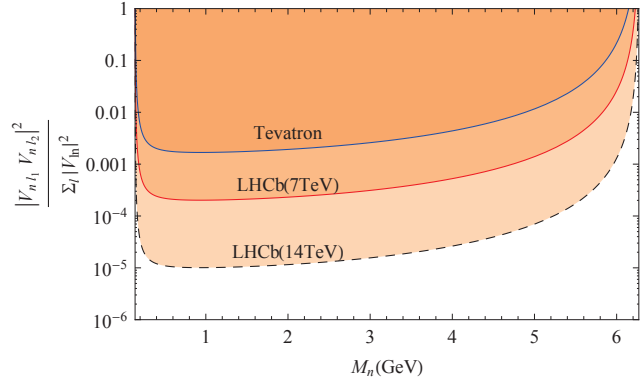}
  \caption{The solid lines stand the low bound of the parameters that is excluded by Tevatron and LHCb through $B_c^\pm\to l_1^{\pm}l_2^{\pm}\pi^\mp$ at 95\% C.L..  And the dashed line is expected to be excluded by LHCb in future. }\label{fig:bound}
\end{center}
\end{figure}

As shown in Ref\cite{Chang:2003cr}, the corresponding cross-section is about $3nb$ at Tevatron($\sqrt{s}=1.96TeV$) and $50nb$ at LHCb($\sqrt{s}=14TeV$). As the desired Luminosity of LHCb is about $10fb^{-1}$, the expect event numbers are  also shown in Fig.(\ref{fig:threshold}).  The LHC has run at 7TeV with total Luminosity $1.11fb^{-1}$ and the production cross-section of $B_c$ is about $22nb$. The exclude region has been shown in Fig.(\ref{fig:bound}) at 95\% C.L..

%The LHC is running at 8TeV, Now the total luminosity is about , the total cross-section $pp\to B^\pm_C+X\to l^\pm l^\pm\pi^\mp +X$ is about $228\times 10^{-4}nb\times \frac{|V_{1n}V_{2n}|^2}{\sum_l|V_ln|^2}$

%%%%%%%%%%%%%%%%%%%%%%%%%%%%%%%%%%%%%%%%%%
\section{Summary}
The neutrino experiments indicate very small neutrino mass. A most natural explanation to such small mass is see-saw mechanism.
% which predicts lepton number violation due to Majorana neutrinos.
In this work, we study the $\Delta L=2$ semi-leptonic decays of $B_c$ meson mediated by Majorana neutrinos and  investigate the contributions  from annihilation and emission diagrams for different Majorana neutrino  mass. The light-cone functions of the mesons are applied to calculate the hadronic matrix element in emission diagrams.  It is found that  the corresponding decay widths are sensitive to the Majorana neutrino mass and the mixing angles.
For a sterile neutrino with mass $m_\pi<m_n<m_B$, the leptonic number violating decay rates of $B_c$ can be enhanced by the annihilation diagrams and may be detectable at LHCb.

%%%%%%%%%%%%%%%%%%%%%%%%%%%%%%%%%%%%%%%%%%%%%
\begin{acknowledgments}
The authors would like to thank Prof. S. Y. Li for his
critical discussions. This work was supported in part by the National Science Foundation of China (NSFC), China Postdoctoral Science Foundation (CPSF) and Natural Science Foundation of Shandong Province.
\end{acknowledgments}

%%%%%%%%%%%%%%%%%%%%%%%%%%%%%%%%%%%%%%%%%%

\end{document}